\documentstyle[floats,prd,aps,epsfig,eqsecnum,12pt]{revtex}
\makeatletter
\newbox\tempboxa
\newdimen\captionboxsubcount 
\def\capsize#1{\captionboxsubcount=#1pt}
\newdimen\captionboxsub
\captionboxsub=\hsize \advance\captionboxsub by -\captionboxsubcount
\advance\captionboxsub by -\captionboxsubcount
\long\def\@makecaption#1#2{
 \setbox\@tempboxa\hbox{#1 #2}
 \ifdim \wd\@tempboxa >\captionboxsub 
\rightskip=\captionboxsubcount \leftskip=\captionboxsubcount #1 #2 
\else \hbox to\hsize{\hfil\box\@tempboxa\hfil} 
 \fi}
\makeatother
\capsize{30}

\begin{document}
\bibliographystyle{unsrt}
\begin{titlepage}
\begin{flushright}
\begin{minipage}{5cm}
\begin{flushleft}
\small
\baselineskip = 13pt
SU-4240-706\\
\end{flushleft}
\end{minipage}
\end{flushright}
\begin{center}
\Large\bf
Mechanism for a next-to-lowest lying scalar meson nonet
\end{center}
\vfil
\footnotesep = 12pt
\begin{center}
\large
Deirdre {\sc Black}
\footnote{Electronic address: {\tt black@physics.syr.edu}}
\quad\quad Amir H. {\sc Fariborz}\footnote{Electronic  
address: {\tt amir@suhep.phy.syr.edu}}
\quad\quad
Joseph {\sc Schechter}\footnote{
Electronic address : {\tt schechte@suhep.phy.syr.edu}}\\
{\it 
\qquad  Department of Physics, Syracuse University, 
Syracuse, NY 13244-1130, USA.} \\
\end{center}
\vfill
\begin{center}
\bf
Abstract
\end{center}
\begin{abstract}
\baselineskip = 17pt
Recent work suggests the existence of a non-conventional lowest-lying scalar
nonet containing the $a_0(980)$.  Then the $a_0(1450)$ and also the
$K_0^*(1430)$ are likely candidates to belong to a conventional p-wave
$q \bar q$ nonet.  However a comparison of their properties with those
expected on this basis reveals a number of puzzling features. It is pointed
out that these puzzles can be resolved in a natural and robust way by
assuming a ``bare'' conventional p-wave scalar $q \bar q$ nonet to mix with
a lighter four quark $qq \bar q \bar q$ scalar nonet to form new
``physical'' states.  The essential mechanism is driven by the fact that
the isospinor is lighter than the isovector in the unmixed $qq \bar q \bar
q$ multiplet. 
\end{abstract}
\begin{flushleft}
\footnotesize
PACS number(s): 13.75.Lb, 11.15.Pg, 11.80.Et, 12.39.Fe 
\end{flushleft}
\vfill
\end{titlepage}
\setcounter{footnote}{0}
\section{Introduction}

The identification and interpretation of the low lying scalar mesons are
questions of great current interest.  A variety of approaches and models
have been explored \cite{Sannino}-\cite{Burakovsky}.  In the effective chiral Lagrangian
approach from which this paper is motivated, a light isoscalar
$\sigma\left( 560 \right)$ in addition to the known light isoscalar
$f_0(980)$ are needed \cite{Sannino} to produce a $\pi \pi$ scattering
amplitude which agrees with experiment.  Similarly a light strange $\kappa
\left( 900 \right)$ state is needed \cite{Black1} to understand the
experimental $\pi K$ amplitude.  These three particles were postulated
\cite{Black2} to form a nonet, taken together with the known isovector
$a_0(980)$.  Consistency of this picture with the properties of the
$a_0(980)$ as seen in ${\eta}^{\prime} \rightarrow \eta\pi\pi$ decay \cite{Fariborz} and as
required in $\pi \eta$ scattering \cite{Black3} was checked .  The pattern
of masses, coupling constants and especially the isoscalar mixing angle was
observed \cite{Black2} to be much closer to the one expected from a
four-quark ($qq\bar {q} \bar {q}$) picture rather than from the conventional
two-quark ($q\bar q$) picture for this scalar nonet.  The four-quark picture
was first proposed by Jaffe \cite{Jaffe} in the framework of the MIT bag
model.  Very recent experiments \cite{VEPP}  on the radiative decays $\phi
\rightarrow \pi\eta\gamma$ and $\phi\rightarrow \pi\pi\gamma$ have been
interpreted \cite{Achasov,Close} as evidence in favor of the four quark picture of
the low-lying scalars $a_0(980)$ and $f_0(980)$.  

Now if one adopts the above picture or, as a matter of fact, any other
picture in which an unconventional non-$q\bar q$ nonet made of the $\sigma
\left( 560 \right)$, $\kappa (900)$, $a_0(980)$ and $f_0(980)$ exists,
there is an interesting puzzle concerning the conventional $q\bar q$ scalar
nonet.  Such a nonet has an interpretation in the constituent quark model as
a p-wave excitation and should therefore share many characteristics of the
other p-wave states (the tensor nonet and two axial vector nonets with
different charge conjugation properties).  To see the puzzling features let
us focus attention on the experimental scalar candidates with non-trivial
isospin quantum numbers in the greater than $1$ GeV energy range.  These
are the isovector $a_0(1450)$ and the strange isospinor $K_0^*(1430)$.
According to the Particle Data Group survey \cite{PDG} (see Table 13.2 on page 110), they are the likely
candidates for a $q\bar q$ scalar nonet.  Then one has the following
unusual features:

  i) The mass of the $a_0(1450)$ (presumably $a_0^+ \sim u\bar d$) is
listed as $1474 \pm 19$ MeV, about $50$ MeV heavier than the strange
$K_0^*(1430)$ (presumably $K_0^{*+} \sim u\bar s$) which has a listed
mass of $1429 \pm 6$ MeV.  Our normal expectation is that the replacement
of the $\bar d$ quark in a $u\bar d$ composite by an $\bar s$ quark should
make the resulting state heavier rather than lighter!  

 ii) On comparison with the corresponding members of the p-wave
$\rm{J^{PC}} = 2^{++}$ nonet, we see that the $q\bar q$ scalar meson
candidates are not lighter; specifically $m[a_0(1474\pm 19)] > m[a_2(1318.1
\pm 0.7)]$ and $m[K_0^*(1429 \pm 6)] \approx m[K_2^*(1432.3 \pm 1.3)]$.
Usually it is expected in the constituent quark model that $\bf{L\cdot S}$
forces should make the spin 0 particle lighter than the corresponding spin
2 particle!  This is experimentally evident in the (perhaps too simple) $c
\bar c$ system where $ m[\chi_{c2}(1P)] = 3556.17 \pm 0.13$ MeV and
$m[\chi_{c0}(1P)] = 3415.1 \pm 0.1$ MeV. 

iii) If $a_0(1450)$ and $K_0^*(1430)$ belong to a conventional nonet their
decay widths into pseudoscalars should be related.  Now, only decay modes
into two pseudoscalars have been observed for these particles:
$K_0^*(1430) \rightarrow \pi K$ and $a_0(1450) \rightarrow \pi \eta, K \bar
K$ and $\pi {\eta}^{\prime}$.  As we will see later, SU(3) symmetry
predicts
\begin{equation}
\Gamma \left[ a_0 \left( 1450 \right)\right] = 1.51 \Gamma \left[ K_0^*
\left( 1430 \right)\right],
\end{equation}
assuming that the total widths are saturated by the decay modes mentioned.
On the other hand the experimental result is 
\begin{equation}
\Gamma \left[ a_0 \left( 1450 \right)\right] = \left( 0.92 \pm 0.12 \right) \Gamma \left[ K_0^*
\left( 1430 \right)\right],
\label{exptl_ratio}
\end{equation}
which clearly differs from the SU(3) prediction.

\begin{figure}
\centering
\epsfig{file=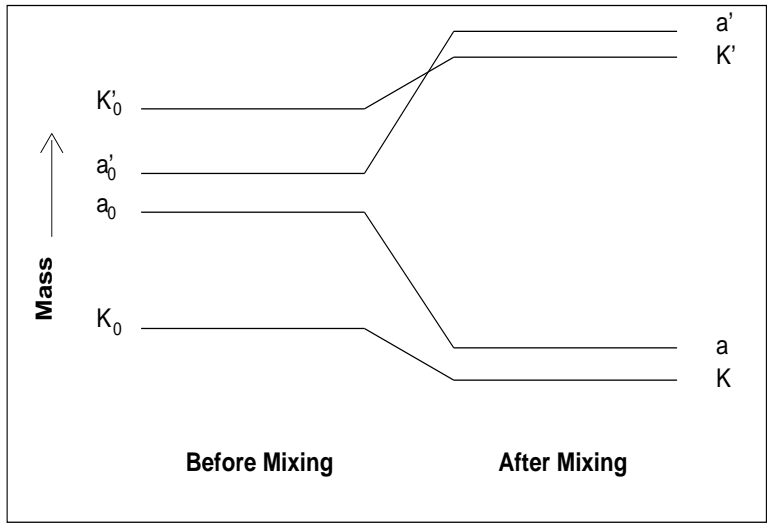, height=3in,width=4in}
\caption
{Mixing of two nonets - $a^{\prime}$, $K^{\prime}$, $a$ and $K$ stand
respectively for the ``physical'' states $a_0(1450)$, $K_0^*(1430)$,
$a_0(980)$ and $\kappa (900)$.  $K_0$ and $a_0$ are the unmixed isospinor
and isovector $qq \bar q \bar q$ states, while $K_0^{\prime}$ and
$a_0^{\prime}$ are the corresponding unmixed $q\bar q$ states.}
\label{perturbation}
\end{figure}

In this note we will show that there exists a model which {\it{naturally}}
provides a solution to these three problems.  This model simply consists of
allowing the $q\bar q$ nonet to mix with a lighter $qq\bar q \bar q$ nonet.
Notice that the isovector in the lighter nonet has a structure $u \bar d s
\bar s$, with two strange quarks.  On the other hand the isospinor in the
lighter nonet has a structure $u \bar s d\bar d$, with only one strange
quark.  Thus, before mixing the lighter nonet will have the isovector
{\it{heavier}} than the strange isospinor.  This situation is illustrated in
Fig. \ref{perturbation}, where the notation is explained.  Details will be
given later, but we can easily see how the scheme works.  The two
isovectors mix with each other as do the two isospinors.  Since the mixing of
the two levels repels them this explains point (ii), why the $q\bar q$
scalars appear heavier than expected.  Similarly the $qq \bar q \bar q$
scalars are pushed down in mass.   Point (i), the level crossing of the
$q\bar q$ isovector and isospinor can be simply understood in the
perturbation theory approximation:  since the $a_0 - a_0^{\prime}$
splitting is smaller than the $K_0 - K_0^{\prime}$ splitting the ``energy
denominator'' for the isovector mixing will be smaller than the one for the
isospinor mixing.  Hence the isovectors will be more strongly repelled.  We
must assume that the $a_0 - K_0$ splitting is large enough so that there is
no level crossing for the lower mass scalars.  Finally point (iii), the
difference in coupling constants of the $K_0^*(1430)$ and the $a_0(1450)$,
can be readily understood from the greater ``contamination'' of the
$a_0(1450)$ wave-function with the four-quark iso-vector state.

In our present work we do not discuss the isoscalars of the scalar $q\bar
q$ nonet.  The reason is that the experimental situation is rather fluid at
the moment, with many candidates.  These include the $f_0(1370)$ (which may
actually correspond to two different states), the $f_0(1500)$ (which may be
a glueball state) and the $f_J(1710)$.  The present scheme suggests a
five-fold mixing between the $\sigma (560)$, the $f_0(980)$, two heavier
$q\bar q$ isoscalar scalars and a glueball.  

\section{Mixing formalism and mass spectra}

Our interest is in investigating the mass spectra and later
the decay properties of scalar mesons which are a mixture of
``conventional'' $q \bar q$ p-wave states and $qq \bar q \bar q$ states.  

In the quark model the usual $q \bar q$ type scalars are grouped into a
nonet, say $N^{\prime}$, with
\begin{equation}
{N^{\prime}}_a^b \sim q_a {\bar q}^b,
\label{conventional}
\end{equation}
where $a$ and $b$ are flavor indices and $q_1, q_2, q_3 = u,d,s$.  So, for
examples, ${N^{\prime}}_3^3$ contains one strange quark and one antistrange quark,
${N^{\prime}}_3^1$ and ${N^{\prime}}_3^2$ contain one strange quark and one light antiquark while ${N^{\prime}}_1^1$ and
${N^{\prime}}_2^2$ have no strange quarks.   
One can also construct ``multiquark'' hadrons, an idea originally discussed
by Jaffe \cite{Jaffe}.  In this paper we focus on a scalar flavor SU(3)
nonet ${N}$ of color SU(3) singlet states with quark composition $qq
\bar q \bar q$.  Within the context of the MIT bag model, Jaffe showed
moreover that the lightest such scalar nonet $N$ should have a
mass less than or in the vicinity of $1$ GeV due to the strong binding
energy of the $qq\bar q \bar q$ configuration arising from hyperfine
interactions between the quarks.  The 4-quark scalar nonet $N$, which
transforms in an identical manner to ${{N}}^{\prime}$ under flavor
SU(3), can naturally be decomposed (this discussion is a summary of
Section II of \cite{Black2}) in terms of ``dual'' flavor quarks (actually
diquarks):
\begin{equation}
N_a^b \sim {T_a}{{\bar T}^b},
\end{equation}
where
\begin{equation}
T_a = \epsilon _{abc} {\bar q}^b {\bar q}^c , \quad \quad {\bar T}^a =
\epsilon^{abc}q_{b}q_{c}.
\label{dual-quarks}
\end{equation} 
So ${N}_3^3$ contains {\it{no}} strange quarks,
${N}_1^3$ and ${N}_2^3$ contain one strange anti-quark each, while
${N}_1^1$, ${N}_2^2$ and  ${N}_1^2$ contain two strange
constituents each.  As explained in the introduction we are not including
the experimentally ambiguous isoscalars in our present discussion and so the pure $q\bar q$ states in
${{N}}^{\prime}$ of interest are the isovector and isospinor; their
charged components are (using the notation of Fig. \ref{perturbation}) :
\begin{equation}
{a^{\prime}_0}^+ \sim u \bar d, \quad \quad {K^{\prime}_0}^+ \sim u \bar s
\end{equation}
and the corresponding members of the $qq \bar q \bar q$ nonet $N$ are 
\begin{equation}
{a_0}^+ \sim us \bar d \bar s, \quad \quad {K_0}^+ \sim ud \bar s \bar d.
\end{equation}
By simply considering the strange quark content of these states, and also
bearing in mind that the $q \bar q$ nonet $N^{\prime}$ presumably lies
in the same mass range as the p-wave axial and tensor meson nonets whereas the
bag-model indication is that the $qq \bar q \bar q$ nonet $N$ should be
less than about $1$ GeV, we expect an ordering of the masses of these states 
\begin{equation}
m_{K_0} < m_{a_0} \le m_{a_0^{\prime}} <  m_{K_0^{\prime}},
\label{ordering}
\end{equation}
as illustrated in Fig. \ref{perturbation}.
Suppose initially that the scalar meson nonet $N^{\prime}$
is ``ideally mixed'' according to the classic idea of Okubo \cite{Okubo}, applied originally to the lightest vector mesons.  ``Ideal mixing''
within the nonet may be defined by the following mass terms of an
effective Lagrangian density for the $q \bar q$ scalars:
\begin{equation}
{\cal L}_{mass}^{\prime} = -a^{\prime} {\rm {Tr}}(N^{\prime}N^{\prime}) -
b^{\prime} {\rm {Tr}}(N^{\prime}N^{\prime}{\cal M}).
\label{mass-Lag-prime}
\end{equation} 
In fact as discussed in \cite{Black2} we may define a generalized ideal
mixing model for the $qq \bar q \bar q$ nonet $N$ by the mass terms:
\begin{equation}
{\cal L}_{mass} = -a {\rm {Tr}}(NN) - b {\rm {Tr}}(NN{\cal M}).
\label{mass-Lag}
\end{equation}
Here ${\cal M}$ is the ``spurion matrix'' [${\cal M}={\rm diag}(1,1,x)$ where $x$ is the ratio of strange to non-strange quark
masses in the usual interpretation].  It
is worth remarking that although (\ref{mass-Lag-prime}) and
(\ref{mass-Lag}) are similar in appearance, the difference at the quark
level between $N$ and $N^{\prime}$ manifests itself through
opposite signs of $b$ and $b^{\prime}$.  This can be seen by noting
that{\footnote{Squared masses are being used since we are working in an
effective Lagrangian framework.  This feature does not play any special
role.}  
\begin{eqnarray}
{m_{K_0}}^2 - {m_{a_0}}^2 &=&   \left( x - 1 \right) b, \\
\nonumber 
{m_{K_0^{\prime}}}^2 - {m_{a_0^{\prime}}}^2 &=& \left( x-1 \right) b^{\prime},
\end{eqnarray}
where the numerical value of $x$ is around 20.5 \cite{Harada-Schechter}.
Here $b^{\prime}$ is taken positive while $b$ is taken negative; this agrees
with counting the number of constituent strange quarks.  

To see whether a mixing between the nonets
$N^{\prime}$ and $N$
can give states whose properties reproduce those of the
experimental scalar isovector and isospinor candidates above $1$ GeV, we consider the simplest invariant term which will induce mixing between $N$ and $N^{\prime}$,
namely 
\begin{equation}
{\cal L}_{mass}^1 =-\gamma {\rm {Tr}} \left( NN^{\prime} \right).
\label{mixing_mass}
\end{equation}

For orientation, we first consider the mixing from the point of view of simple perturbation theory applied to two two-state systems.  For the
isovectors and isospinors we have the $2 \times 2$ mixing matrices 
\begin{equation}
M_a^2 =
 \left[ \begin{array}{c c}
m_{a_0}^2 & \gamma \\
\gamma & m_{a_0^{\prime}}^2 \end{array} \right] \quad {\rm{and}} \quad M_K^2 =  
 \left[ \begin{array}{c c}
m_{K_0}^2 & \gamma \\
\gamma & m_{K_0^{\prime}}^2 \end{array} \right]. 
\label{mixing_matrices}
\end{equation}   

At second order in perturbation theory we see that the shifts in the
square masses for the $a_0 - a_0^{\prime}$ and $K_0 - K_0^{\prime}$ systems
have magnitudes 
\begin{equation}
{\Delta}_a = \frac {\gamma^2}{m_{a_0^{\prime}}^2 - m_{a_0}^2} \quad {\rm{and}} \quad 
{\Delta}_K = \frac {\gamma^2}{m_{K_0^{\prime}}^2 - m_{K_0}^2},
\end{equation}
respectively.  Clearly the ordering of the masses in Eq. (\ref{ordering})
implies that ${\Delta}_a > {\Delta}_K$ and so if the gap between $m_{a_0}$
and $m_{K_0}$ is sufficiently large relative to the $a_0^{\prime}-K_0^{\prime}$ mass difference we will naturally be able to obtain the
level-crossing behavior of Fig. 1.

Next we proceed to an exact treatment.  Invariance of the trace of
$M_a^2 $ upon diagonalization implies that 
\begin{equation}
m_{a_0}^2 + m_{a_0^{\prime}}^2 =  m_{a}^2 + m_{a^{\prime}}^2
\label{trace}
\end{equation}
with an analogous equation holding for the isospinors.  Using this condition and the eigenvalue equation
\begin{equation}
\left( m_{a_0}^2 - m_a^2 \right) \left( {m_{a_0^{\prime}}}^2 - m_a^2 \right)
- {\gamma}^2 = 0,
\end{equation} 
we solve for the masses of the original unmixed states to obtain:
\begin{equation}
{m_{{a_0}/{a_0^{\prime}}}}^2 = \frac{1}{2} \left[ m_a^2 + {m_{a^{\prime}}}^2
\mp \sqrt{ {\left( {m_{a^{\prime}}}^2 - m_a^2 \right)}^2 - 4{\gamma}^2} \right].
\end{equation}
Analogous equations follow from the diagonalization of $M_K^2$.  These
equations may be read as giving for each value ${\gamma}^2$ the
corresponding masses of the unmixed states which will, upon inclusion of ${\cal
L}_{mass}^1$, lead to the experimentally known physical masses.  Reality of the masses implies that $4{\gamma}^2 \leq {\rm{Min}} \{ {\left(
{m_{a^{\prime}}}^2 - m_a^2 \right)}^2,{\left( {m_{K^{\prime}}}^2 - m_K^2
\right)}^2 \}$.  

We are interested in a scenario where the ordering of the unmixed masses is
as in Eq. (\ref{ordering}).  We find that this can happen provided that  $m_{a^{\prime}}^2
- m_a^2  <  m_{K^{\prime}}^2 - m_K^2$ (which holds for most of the
experimentally allowed range of masses) because in this case the behavior of the
bare masses is as shown in Fig. 2.  Of course the bare and
physical masses must coincide for $\gamma = 0$.  We define ${{\gamma}_{max}}^2 = \frac{1}{4}{\left(m_{a^{\prime}}^2 - m_a^2
\right)}^2$; the value of ${{\gamma}_{max}}^2$ depends (since $m_a$ is very
accurately known) on the exact value of $m_{a^{\prime}}$.  For $\gamma =
{\gamma}_{max}$ the $I=1$ states are maximally mixed and
the unmixed states are degenerate with square masses equal to $\frac{1}{2} (m_a^2 +
{m_{a^{\prime}}}^2)$.  We see from Fig. 2 that the choice $\gamma =
{\gamma}_{max}$ is expected to result in the largest splitting of the bare $q \bar q$
masses ${m_{K_0^{\prime}}}^2 - {m_{a_0^{\prime}}}^2$.

\begin{figure}
\centering
\epsfig{file=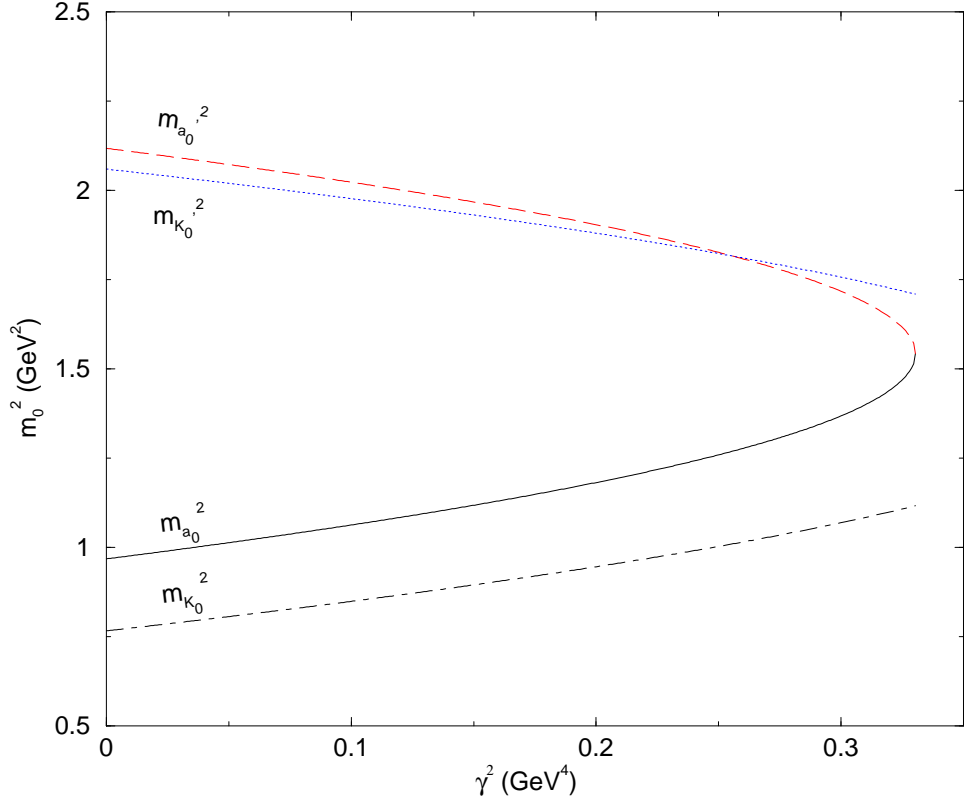, height=5in,angle=270}
\caption
{Evolution, as a function of ${\gamma}^2$, of the bare masses needed to produce
the physical ones.  Of course, the bare and physical masses coincide for
${\gamma}^2=0$.  This picture corresponds to the case ${m_{K^{\prime}}}^2 -
{m_K}^2 > {m_{a^{\prime}}}^2 - {m_a}^2$, which holds for the central
``experimental'' values.  Here the plot is shown for the physical masses
chosen at the end of Section II. The dot-dashed curve is ${m_{K_0}}^2$,
the solid curve is ${m_{a_0}}^2$, the dotted curve is
${m_{K_0^{\prime}}}^2$ and the dashed curve is ${m_{a_0^{\prime}}}^2$.}
\label{masses_fig}
\end{figure}

A detailed numerical search shows that the largest value of this
splitting is in fact obtained for ${\gamma}^2 = {{\gamma}_{max}}^2 = 0.33$
${\rm{GeV}}^4$ and with the choice of
physical masses (within the allowed ``experimental'' range)
\begin{eqnarray}
m_a = 0.9835 \hskip 0.2cm {\rm{GeV}} &,& \quad \quad m_{a^{\prime}} = 1.455
\hskip 0.2cm {\rm{GeV}},  \nonumber \\
m_K = 0.8750 \hskip 0.2cm {\rm{GeV}} &,& \quad \quad m_{K^{\prime}} =
1.435 \hskip 0.2cm  {\rm{GeV}}.
\label{physical_masses}
\end{eqnarray}
Fig. \ref{allowed_points_fig} shows all the allowed points and their
corresponding mass splittings. 
Notice that $m_K$ is obtained from the analysis of $\pi K$ scattering given
in \cite{Black1}.  This then yields the following masses for the unmixed
states:
\begin{eqnarray}
m_{a_0}= m_{a_0^{\prime}} &=& 1.24 \hskip 0.2cm {\rm{GeV}},  \nonumber \\
m_{K_0} = 1.06 \hskip 0.2cm {\rm{GeV}}&,& \quad m_{K_0^{\prime}} = 1.31
\hskip 0.2cm {\rm{GeV}}.
\label{bare_masses}
\end{eqnarray}

We see that $m_{K_0^{\prime}} - m_{a_0^{\prime}} \approx 65$ MeV which is comparable with
the analogous splitting of the tensor and axial families of
order $100$ MeV.  We also notice that in addition to satisfying the
ordering in Eq. (\ref{ordering}) [which can be an explanation for puzzle
(i)], we can understand puzzle (ii) in this picture since the unmixed
$q \bar q$ scalar states are lighter than the analogous tensors.
Specifically, we have that 
 $m[a_0^{\prime}]  < m[a_2(1318.1)]$ and $m[K_0^{\prime}] < m[K_2^*(1432.3 \pm 1.3)]$.

\begin{figure}
\centering
\epsfig{file=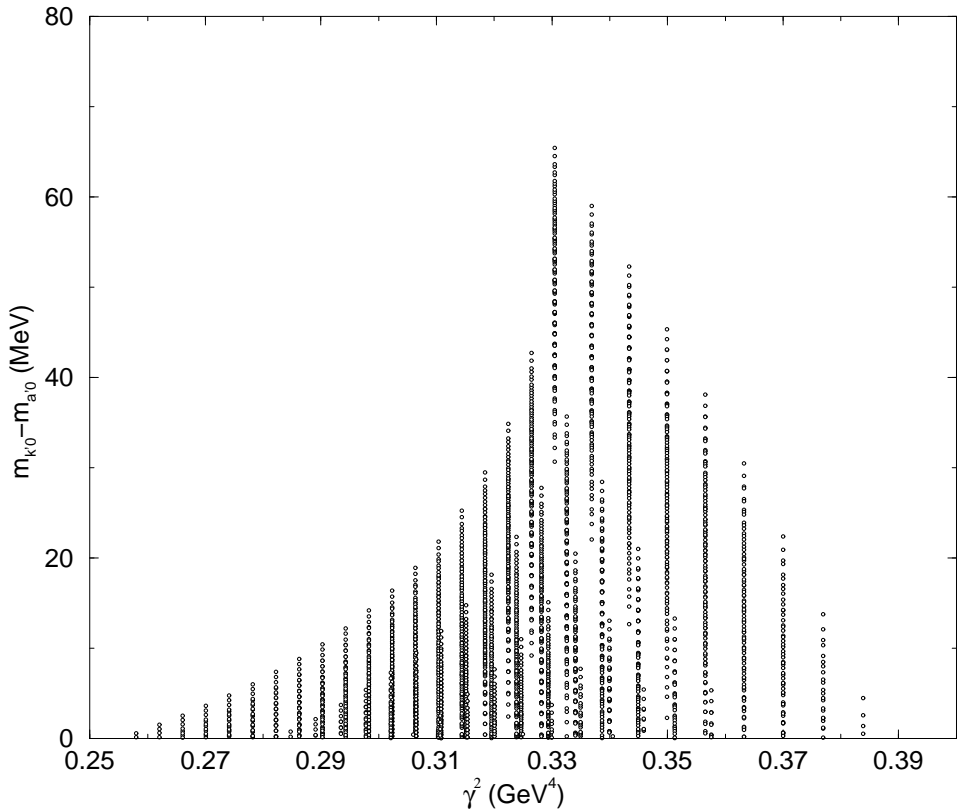, height=5in,angle=270}
\caption
{Scan showing the values of the bare mass splitting $m_{K_0^{\prime}} -
m_{a_0^{\prime}}$ resulting from different experimentally allowed masses of
$a$, $a^{\prime}$, $K$ and $K^{\prime}$ and of ${\gamma}^2$.  The mechanism
gives the correct ordering for the approximate range 
$0.26 < {\gamma}^2 < 0.38 \hskip 0.2cm {\rm{GeV}}^4$.}
\label{allowed_points_fig}
\end{figure}
  
\section{Decay widths}

In this section we address the third puzzle presented in the introduction.
The total widths of the $a_0(1450)$ and the $K_0^*(1430)$ are
listed in the Review of Particle Physics \cite{PDG} as:
\begin{equation}
{\Gamma}^{tot} \left[ K_0^* \left( 1430 \right) \right] = 287 \pm 23 \hskip 0.2cm {\rm{MeV}} \quad
{\rm{and}} \quad {\Gamma}^{tot} \left[ a_0 \left(1450 \right) \right] = 265
\pm 13 \hskip 0.2cm {\rm{MeV}},
\label{exptl_widths}
\end{equation}
which implies the ratio in Eq. (\ref{exptl_ratio}).  The only listed decay
mode of the $K_0^*(1430)$ is $\pi K$ with a branching fraction of $\left(
93 \pm 10 \right) \%$ which is close to $100 \%$.  On the other hand for the
$a_0(1450)$, the experimental knowledge of the exclusive decay modes
is less certain; the $\pi \eta$, $K \bar K$ and $\pi
{\eta}^{\prime}$ modes are listed as ``seen'' without stating any branching
fractions.  In the detailed listings the following ratios are presented:
\begin{equation}
\frac{{\Gamma} \left[ a_0 \left( 1450 \right) \rightarrow K \bar K \right]}{{\Gamma} \left[
a_0 \left(1450 \right) \rightarrow \pi \eta \right]} = 0.88 \pm 0.23 \quad {\rm{and}} \quad
\frac{{\Gamma} \left[ a_0\left(1450 \right) \rightarrow \pi {\eta}^{\prime} \right]}{{\Gamma} \left[
a_0\left(1450 \right) \rightarrow \pi \eta \right]} = 0.35 \pm 0.16.
\label{exptl_partial_widths}
\end{equation}
In this section we also denote the physical state $a^{\prime}=a_0(1450)$ by
$a_0^*$ and the physical state 
$K^{\prime}=K_0^*(1430)$ by $K_0^*$.  Despite the uncertainty, for the purpose of our
analysis we shall assume that the $\pi \eta$, $ K \bar K$ and $\pi
{\eta}^{\prime}$ modes saturate the $a_0(1450)$ decays and that their
ratios expressed above hold as stated.  

Using isoptopic spin invariance the scalar-pseudoscalar-pseudoscalar
trilinear interaction terms relevant for these decay channels can be
written as{\footnote{Derivative coupling is being used because we want our
Lagrangian to be a piece of a chiral invariant object.  See Appendix B of
\cite{Black2}.}  
\begin{eqnarray}
-{\cal L} & = & \frac{\gamma_{a_0^* K  K}}{\sqrt 2} \partial_\mu {\bar
K} \mbox{\boldmath ${\tau}$} \cdot {\bf a_0^*} \partial_\mu {K} +  \gamma_{a_0^* \pi\eta} {\bf a_0^*} \cdot \partial_\mu \mbox{\boldmath ${\pi}$} \partial_\mu \eta + 
 \gamma_{a_0^* \pi\eta'} {\bf a_0^*} \cdot \partial_\mu \mbox{\boldmath
${\pi}$} \partial_\mu \eta'  \nonumber \\ 
& + & \frac{\gamma_{K_0^* K \pi}}{\sqrt 2} \left(
\partial_\mu {\bar K} \mbox{\boldmath ${\tau}$} \cdot \partial_\mu {\mbox{\boldmath ${\pi}$}}
{K_0^*} + h.c. \right).
\label{Lagrangian}
\end{eqnarray}
Hence the perturbative decay width of the $K_0^*(1430)$ is
\begin{equation}
\Gamma \left( K_0^* \rightarrow \pi K \right) = \frac{3}{2} \frac
{{\gamma_{K_0^* K \pi}}^2}{32 \pi} \frac{q}{{m_{K_0^*}}^2} {\left(
{m_{K_0^*}}^2 - {m_{\pi}}^2 - {m_K}^2 \right)}^2,
\end{equation}
where $q$ is the center of mass momentum of the decay products.  Analogous expressions follow for the $a_0(1450)$ partial widths.  Thus we have that 
\begin{equation}
\begin{array}{c c}
{\Gamma} \left(
a_0^* \rightarrow \pi \eta \right) = 0.0099 {{\gamma}_{a_0^* \pi\eta}}^2, \quad & \quad {\Gamma} \left(a_0^* \rightarrow \pi {\eta}^{\prime} \right) = 0.0028 {{\gamma}_{a_0^* \pi{\eta}^{\prime}}}^2, \\ {\Gamma} \left(
a_0^* \rightarrow K \bar K \right) = 0.0070 {{\gamma}_{a_0^* K \bar K}}^2,\quad & \quad {\Gamma} \left(
K_0^* \rightarrow \pi K \right) = 0.0143 {{\gamma}_{K_0^* \pi K}}^2.
\end{array}
\label{theor_widths}
\end{equation} 

Let us initially suppose that the $a_0(1450)$ and $K_0^*(1430)$ are members
of a hypothetical unmixed scalar $q \bar q$ nonet $N^{\prime}$,
i.e. $\gamma = 0$.  Then their
decays into two pseudoscalars are presumably described by the interaction
\begin{equation}
{\cal L}_{{{N}}^{\prime}\phi \phi} = 2{A}^{\prime} {\rm Tr} \left(
N^{\prime}{\partial_\mu}{\phi}{\partial_\mu}{\phi}\right),
\label{SU3Lag}
\end{equation}
where ${\phi}_a^b$ is the matrix of pseudoscalar fields.  This pure $q \bar q$
coupling term, when expanded into individual fields and compared with (\ref{Lagrangian}) above (where the coupling constant conventions are defined), leads to the identifications:
\begin{equation} 
{\gamma}_{a_0^* \pi\eta} = - 2 \sqrt{2} {\rm{cos}}{\theta}_p  A^{\prime},
\quad 
{\gamma}_{a_0^* \pi{\eta}^{\prime}} = - 2 \sqrt{2} {\rm{sin}}{\theta}_p 
A^{\prime}, \quad 
{\gamma}_{a_0^* K \bar K} = {\gamma}_{K_0^* \pi K} = - 2 A^{\prime},
\label{SU3couplings}
\end{equation}
where ${\theta}_p$ is the pseudoscalar mixing angle, which we take to be
$37^{\circ}$ \cite{Harada-Schechter}.  Now if we substitute into (\ref{theor_widths}) we find the
$q \bar q$ SU(3) predictions for the ratios of the total widths:
\begin{equation}
\frac{{\Gamma}^{tot} \left( a_0^* \right)}{{\Gamma} \left( K_0^*
\rightarrow  \pi K \right)} = 1.51, 
\label{SU3_1}
\end{equation}
and for the partial $a_0(1450)$ widths: 
\begin{equation}
\frac{{\Gamma} \left( a_0^* \rightarrow K \bar K \right)}{{\Gamma} \left(
a_0^* \rightarrow \pi \eta \right)} = 0.55, \quad
\frac{{\Gamma} \left( a_0^* \rightarrow \pi {\eta}^{\prime} \right)}{{\Gamma} \left(
a_0^* \rightarrow \pi \eta \right)} = 0.16.
\label{SU3_2}
\end{equation}

We see that while (\ref{SU3_2}) are just a little below the experimentally
allowed ratios (\ref{exptl_partial_widths}), the ratio (\ref{SU3_1}) is not
consistent with the experimental ratio which follows from
(\ref{exptl_widths}).  Thus considering the $a_0(1450)$ and $K_0^*(1430)$
as members of a pure $q\bar q$ SU(3) nonet does not give good agreement with
experiment.

Next we study the predictions for the decay widths of the $a_0(1450)$ and
the $K_0^*(1430)$ in the mixing picture of Section II.  In \cite{Black2} we
discussed the general SU(3) flavor invariant coupling of members of a
scalar nonet to two pseudoscalars.  For the case of the $q\bar q$ scalar nonet $N^{\prime}$ the most standard form is as taken in (\ref{SU3Lag}) above.  However, for the 4-quark nonet, $N$, a more natural structure which to a first approximation reproduce
s the scalar decay pattern is:
\begin{equation}
{\cal L}_{N\phi \phi} =
A{\epsilon}^{abc}{\epsilon}_{def}N_{a}^{d}{\partial_\mu}{\phi}_{b}^{e}{\partial_\mu}{\phi}_{c}^{f}.
\label{multiquarkLag}
\end{equation} 
For $qq \bar q \bar q$ mesons it seems reasonable that the dominant decays
will simply be ones that involve a ``falling-apart'' \cite{Jaffe}, or
rearrangement, of the 4 quarks into two $q\bar q$ mesons.  So for example
since ${N}_3^3$ contains no strange quarks one would expect its
decay into $K \bar K$ to be suppressed.  Indeed the Lagrangian
(\ref{multiquarkLag}) predicts zero coupling of $N_3^3$ into $K \bar K$. 

Upon diagonalization of (\ref{mixing_matrices}) the physical isospinors are
$K= \kappa (900)$ and $K^{\prime} = K_0^*(1430)$ and we take the mixing convention:
\begin{equation}
\left( \begin{array}{c} \kappa \left( 900 \right) \\ K_0^* \left( 1430 \right)  \end{array} \right) = \left(
\begin{array}{c c} {\rm cos} \psi_K & -{\rm sin} \psi_K \\ {\rm sin}
\psi_K & {\rm cos} \psi_K \end{array} \right) \left( \begin{array}{c}
K_0 \\  K_0^{\prime} \end{array} \right).
\label{mixing-convention_K}
\end{equation}
Likewise the isovector mass eigenstates are $a=a_0(980)$ and $a^{\prime}=a_0(1450)$ given by 
\begin{equation}
\left( \begin{array}{c} a_0 \left( 980 \right) \\ a_0 \left( 1450 \right)  \end{array} \right) = \left(
\begin{array}{c c} {\rm cos} \psi_a & -{\rm sin} \psi_a \\ {\rm sin}
\psi_a & {\rm cos} \psi_a \end{array} \right) \left( \begin{array}{c}
a_0 \\  a_0^{\prime} \end{array} \right),
\label{mixing-convention_a}
\end{equation}
where the mixing angles are obtained as:
\begin{equation}
{\rm{tan}} \left( 2\psi_K \right) = \frac{2\gamma}{{m_{K_0^{\prime}}}^2 - {m_{K_0}}^2} \quad {\rm{and}} \quad {\rm{tan}} \left( 2\psi_a \right) = \frac{2\gamma}{{m_{a_0^{\prime}}}^2 - {m_{a_0}}^2}.
\label{mixing_angles}
\end{equation}
Now if we take the total trilinear interaction Lagrangian density to be the
sum of (\ref{SU3Lag}) and (\ref{multiquarkLag}) and expand the relevant
unmixed isovector and isospinor members of $N$ and $N^{\prime}$ in
terms of the physical fields using the mixing convention above, we find
that [see (\ref{Lagrangian}) and compare with the unmixed case (\ref{SU3couplings})]
\begin{eqnarray} 
{\gamma}_{a_0^* \pi\eta} &=& - 2 \left( {\rm{sin}}{\psi}_a
{\rm{sin}}{\theta}_p {\rm{A}} + \sqrt{2} {\rm{cos}}{\theta}_p
{\rm{cos}}{\psi}_a  A^{\prime}\right), \nonumber \\   {\gamma}_{a_0^*
\pi{\eta}^{\prime}} &=& 2 \left( {\rm{sin}}{\psi}_a {\rm{cos}}{\theta}_p
{\rm{A}} - \sqrt{2} {\rm{sin}}{\theta}_p  {\rm{cos}}{\psi}_a 
A^{\prime}\right), \nonumber \\ 
{\gamma}_{a_0^* K \bar K} &=&  - 2 \left( {\rm{sin}}{\psi}_a A +
{\rm{cos}}{{\psi}_a}A ^{\prime}\right),  \nonumber \\  {\gamma}_{K_0^* \pi K}
&=& - 2 \left({\rm{sin}}{\psi}_K A +  {\rm{cos}}{{\psi}_K}A ^{\prime}\right).
\label{mixed_couplings}
\end{eqnarray}
Again we calculate the widths using Eqs. (\ref{theor_widths}) and compare their ratios with experiment.  It turns out to be helpful to begin by analyzing these ratios in different regions.  

First we consider the limit where $|{\frac{A^{\prime}}{A}}|$ is large.  In this region,
\begin{equation}
\frac{{\Gamma} \left( K_0^* \rightarrow \pi K \right)}{{\Gamma} \left(
a_0^* \rightarrow \pi \eta \right)} \sim \left[1.444 \frac{1}{2 {\rm{cos}}^2{\theta}_p}\right] \left[ {\frac{{\rm{cos}}{\psi}_K}{{\rm{cos}}{\psi}_a}} \right]^2.
\label{tot_widths_mixed_limit}
\end{equation}
The term in the first bracket is what we obtained above from the couplings in (\ref{SU3couplings}) and so gives the prediction (\ref{SU3_1}) which is smaller than experiment.  Recalling (\ref{mixing_angles}) we see that for mixing angles ${\psi}_a,
{\psi}_K \in \left[ 0,\frac{\pi}{4} \right]$ (which is all that is needed
 in this limit since the relative sign introduced by considering ${\psi}_a, {\psi}_K \in
\left[ - \frac{\pi}{4,}0 \right]$ - and so for the opposite sign of
$\gamma$- may be absorbed in a change of the relative sign of A and ${\rm{A}}^{\prime}$) we will always have that ${\psi}_a > {\psi}_K$ since we are
considering the scheme where the ordering of the masses is as in
Eq. (\ref{ordering}).  Hence the ratio (\ref{tot_widths_mixed_limit})
always {\it{increases}} relative to the $q \bar q$ prediction in this limit and so the ratio of the total widths
(\ref{SU3_1}) will {\it{decrease}}, as required.  This behavior is independent of
the choice of experimental masses as long as they satisfy ${m_{K^{\prime}}}^2 - {m_K}^2
> {m_{a^{\prime}}}^2 - {m_a}^2$, which is true whenever the mechanism
works in order to produce the correct level-crossing behavior for the
masses.  For example, for the illustrative input masses considered at the
end of Section II we have that ${\psi}_a = \frac{\pi}{4}$ and ${\psi}_K
\approx 31^{\circ}$ which implies that 
\begin{equation}
\frac{{\Gamma}\left(a_0^* \rightarrow \pi\eta\right)}{{\Gamma}\left( K_0^* \rightarrow  {\pi K}\right)} \approx 0.606
\end{equation}
giving that 
\begin{equation}
\frac{{\Gamma}^{tot}\left(a_0^*\right)}{{\Gamma}\left(K_0^* \rightarrow \pi K\right)} \approx 1.036.
\label{improvement}
\end{equation}
Within the errors quoted in \cite{PDG} this agrees with the experimental
result (\ref{exptl_widths}) and is much closer than the $q \bar q$ prediction of Eq. (\ref{SU3_1}).
Furthermore from (\ref{mixed_couplings}) for large
$|{\frac{A^{\prime}}{A}}|$ we have the same prediction as in Eq. (\ref{SU3_2}).   

Next we suppose conversely that $|{\frac{A}{{A}^{\prime}}}|$ is large.  In this region,
\begin{equation}
\frac{{\Gamma} \left( K_0^* \rightarrow \pi K \right)}{{\Gamma} \left(
a_0^* \rightarrow \pi \eta \right)} \sim
\left[1.444 \frac{1}{ {\rm{cos}}^2{\theta}_p}\right]
{ \left[ \frac{{\rm{sin}}{\psi}_K}{{\rm{sin}}{\psi}_a} \right]}^2 \approx 2.115,
\label{tot_widths_mixed_limit_b}
\end{equation}
for ${\psi}_K = 31^{\circ}$.  The ratios of the $a_0(1450)$ widths now
become 
\begin{equation}
\frac{{\Gamma} \left( a_0^* \rightarrow K \bar K \right)}{{\Gamma} \left(
a_0^* \rightarrow \pi \eta \right)} \sim 0.7071 \frac{1}{
{\rm{sin}}^2 {\theta}_p}
 = 1.95, \quad \quad
\frac{{\Gamma} \left( a_0^* \rightarrow \pi {\eta}^{\prime} \right)}{{\Gamma} \left(
a_0^* \rightarrow \pi \eta \right)} = 0.2828
{\rm{cot}}^2{\theta}_p = 0.49.
\label{mixed_partial_widths_limit_b}
\end{equation}
In this limit, where it is the $qq\bar{q}\bar{q}$ decay modes of the
$a_0(1450)$ and the $K_0^*(1430)$ 
that dominate, we see that in particular the
first ratio in (\ref{mixed_partial_widths_limit_b}) is well outside the
experimentally allowed range.

\begin{figure}
\centering
\epsfig{file=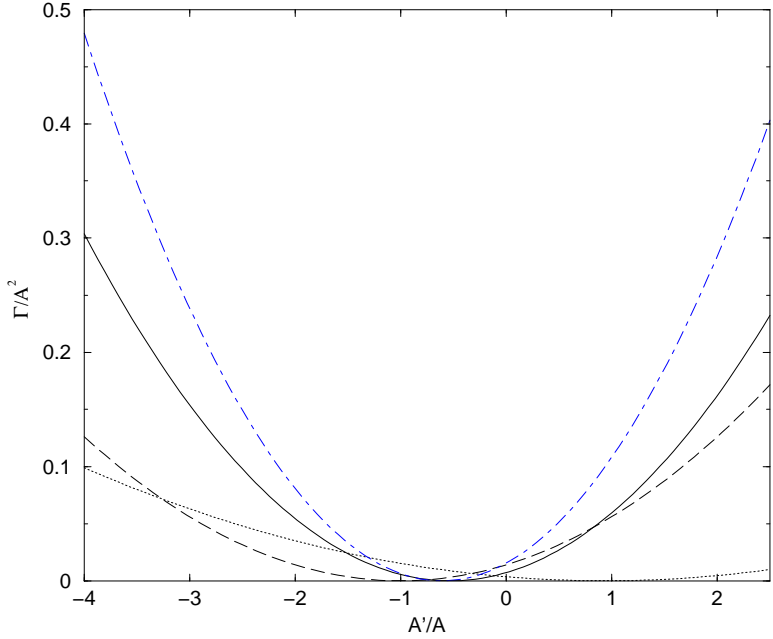, height=4in,angle=270}
\caption
{Plot of $\frac{\Gamma}{{\rm{A}}^2}$ against
$\frac{{\rm{A}}^{\prime}}{\rm{A}}$.  The dot-dashed curve is $\Gamma
\left[ K_0^* \left(1430 \right) \rightarrow \pi K \right]$, the solid
curve is $\Gamma \left[ a_0 \left(1450 \right) \rightarrow  \pi \eta
\right]$, the dashed curve is $\Gamma \left[ a_0 \left(1450
\right) \rightarrow K \bar K \right]$ and the dotted curve is 
$\Gamma \left[ a_0 \left(1450 \right) \rightarrow  \pi {\eta}^{\prime} \right]$.}
\label{widths_fig}
\end{figure}

For ${\rm{A}}^{\prime} \sim {\rm{A}}$ a graphical analysis is helpful
since in this region the ratios of the widths blow up.  In Fig. 3 we plot
the widths themselves (up to an overall normalization of $\frac{1}{A^2}$).
It is seen that, for the central values of (\ref{exptl_widths}) and
(\ref{exptl_partial_widths}) the correct width order ${\Gamma} \left( K_0^*
\rightarrow \pi K \right) > {\Gamma} \left(a_0^* \rightarrow \pi \eta
\right) > {\Gamma} \left( a_0^* \rightarrow K \bar K \right) > {\Gamma}
\left( a_0^* \rightarrow \pi {\eta}^{\prime} \right)$ is obtained for the
``asymptotic'' regions $\frac{{\rm{A}}^{\prime}}{\rm{A}} > 1.2$ and
$\frac{{\rm{A}}^{\prime}}{\rm{A}} < -3.2$.  Inside, where $ -3.2 <
\frac{{\rm{A}}^{\prime}}{\rm{A}} < 1.2$, the correct width order cannot be
obtained for the central values.  

In summary, the above analysis shows that for large
$|\frac{{\rm{A}}^{\prime}}{\rm{A}}|$ the mechanism significantly
improves [see for example (\ref{improvement})] the ratio $\frac{{\Gamma}^{tot}\left(a_0^*\right)}{{\Gamma}\left({K_0^*\rightarrow \pi K}\right)}$
compared with the prediction based on a pure $q\bar q$ description of the
$a_0(1450)$ and the $K_0^*(1430)$.  Outside of this asymptotic region more detailed analysis is
needed and requires additional experimental guidance.  

\section{Numerical Analysis of the Decay Widths}

In this section we give a more detailed numerical analysis of the decay
widths.  We will take into account the experimental uncertainties for
comparison with theory.  Furthermore we will include a more general form
of decay interaction.   Finally the decay widths of the lighter scalars
$K=\kappa(900)$ and $a=a_0(980)$ will also be discussed.  The input masses
will be kept the same as in Eq. (\ref{physical_masses}), and we will continue to use
${\gamma}^2={\gamma_{max}}^2 = 0.33$ ${\rm{GeV}}^4$.  The
general interaction Lagrangian describing the decay widths has
the form (see \cite{Black2})
\begin{eqnarray}
{\cal L}_{int.} &=& 
A \epsilon^{abc}\epsilon_{def}
N_a^d\partial_\mu\phi^e_b\partial_\mu\phi^f_c
+
C {\rm Tr} (N\partial_\mu\phi) {\rm Tr} (\partial_\mu\phi) 
\nonumber \\
&+&
A' \epsilon^{abc}\epsilon_{def}
{N'}_a^d\partial_\mu \phi^e_b \partial_\mu \phi^f_c
+
C' {\rm Tr} (N'\partial_\mu\phi) {\rm Tr} (\partial_\mu\phi)
+\cdots
\label{int_Lag}
\end{eqnarray}
where the three dots stand for terms which do not contribute to isovector
or isospinor decays.

We first consider the limit $C=0$ and $C'=2A'$, as in Section III.
In this limit,  the above Lagrangian
simplifies 
\footnote{It is helpful to use the identity: \\
$A^{\prime} \epsilon^{abc}\epsilon_{def}
{N^{\prime}}_a^d\partial_\mu \phi^e_b \partial_\mu \phi^f_c
= 2 A^{\prime} {\rm Tr} (N^{\prime}\partial_\mu \phi \partial_\mu\phi)
-  A^{\prime} {\rm Tr} (N^{\prime}) {\rm Tr} (\partial_\mu \phi \partial_\mu\phi)
- 2 A^{\prime} {\rm Tr} (N^{\prime}\partial_\mu \phi) {\rm Tr} (\partial_\mu\phi)
+ A^{\prime} {\rm Tr} (N^{\prime}) {\rm Tr} (\partial_\mu \phi){\rm Tr} (\partial_\mu\phi)
$.}
to
\begin{equation}
{\cal L}_{int.} =
A \epsilon^{abc}\epsilon_{def}
N_a^d\partial_\mu\phi^e_b\partial_\mu\phi^f_c
+
2 A' {\rm Tr} (N'\partial_\mu \phi \partial_\mu\phi)
+ \cdots
\label{int_simp}
\end{equation} 
We scan the $AA'$ parameter space
numerically and search for regions consistent with the
available experimental data on the decay widths of these scalars.  
We start with the $a_0(1450)$ decay widths as they impose the strongest 
restrictions on the parameter space. 
First,  we find that the experimental estimate $\Gamma^{tot.}[a_0(1450)] =
265 \pm 13$ MeV restricts $A$ and
$A'$ to the perimeter of the ellipse shown in Fig. \ref{Fig_a0_1450} --
the thickness of the   perimeter
 is related to the 13 MeV uncertainty of the decay width.   We then 
search for regions that are consistent with the
current experimental estimates on the  ratios  
$\Gamma [a_0(1450)\rightarrow K {\bar K} ]/\Gamma [a_0(1450)\rightarrow 
\pi \eta] = 0.88\pm 0.23$, and 
$\Gamma [a_0(1450)\rightarrow \pi\eta' ]/\Gamma [a_0(1450)\rightarrow  
\pi \eta] = 0.35 \pm 0.16$.   Regions consistent with the first
and second ratios are respectively shown by dark and light shading. 
The vertical axis corresponds to the conventional interaction term
for $q{\bar q}$ nonets,
whereas the horizontal axis represents a natural 	
interaction for $qq{\bar
q}{\bar
q}$ nonets, as previously discussed.
We see in  Fig. \ref{Fig_a0_1450} that within our model we cannot
exactly describe the current
experimental data on  the partial decay widths of $a_0(1450)$.
Obviously a natural  four-quark interaction is far from the allowed
regions,  
while a natural  two-quark interaction  seems  to be a favorable  scenario
for 
description of the available experimental data.  
Also small distortions from the natural  two-quark interaction, although
slightly
improving the situation, do not exactly describe the data.  This
is described more quantitatively in Table \ref{T_a0_1450}.  
We have fitted the prediction of our model
for the total decay width as well as the decay ratios,  to the
above  experimental estimates and searched for the best values of $A$
and $A'$.  The natural two-quark
interaction (column one) is compared with the more general case that
natural four-quark interactions are also allowed (column two).  Although
the 
$\chi^2$ of the  fit gets  slightly reduced, effectively the best point
remains  around the  natural two-quark interaction.

In order to see whether we could get a better description of $a_0(1450)$, 
we have also extended our investigations to the more general case where
$C\ne 0$ and $C'\ne 2 A'$, i.e. working with the general Lagrangian
(\ref{int_Lag}). The result 
is given in column three of Table  \ref{T_a0_1450}, indicating that even
with the introduction of more general interaction terms, the
current experimental data is still not exactly described.  We also notice
that
in this case
$C'-2A'\approx 0$ and that $|2 A'|$ dominates $A$ and $C$.  Thus the best
fit in this case also is similar to column I, although the fit is
slightly improved.
Therefore, the simplified model Eq. (\ref{int_simp}) already 
provides a  reasonable picture for understanding the nature of the
$a_0(1450)$. We should emphasize, however,  that the available
experimental 
estimates of the decay channels of $a_0(1450)$ are not very accurately
known.
More accurate experimental data on $a_0(1450)$ would be useful for our
purposes.
\begin{figure}
\centering
\epsfig{file=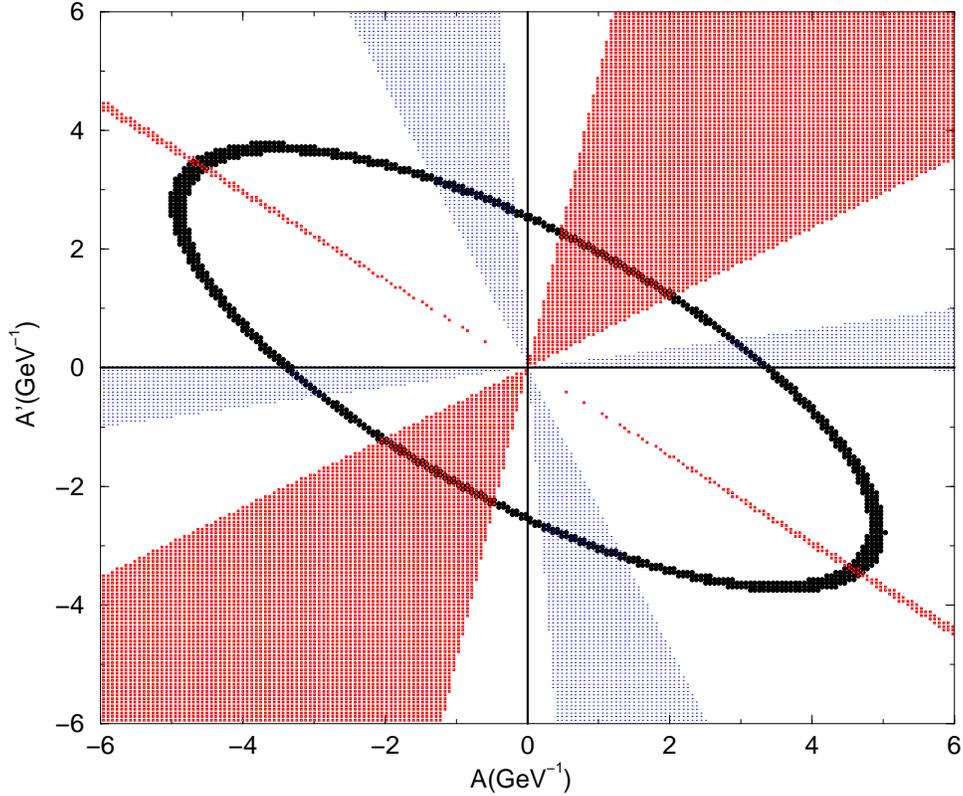, height=5in, angle=270}
\caption
{
Regions in the $AA'$ parameter space [see Eq. (\ref{int_simp})] consistent
with the currently available experimental estimates on the decay widths of $a_0(1450)$.  Points
on the ellipse are consistent with the total decay width of $a_0(1450)$.
Dark and light gray regions respectively represent
points consistent with
the experimental ratio 
$\Gamma [a_0(1450)\rightarrow K {\bar K} ]/\Gamma
[a_0(1450)\rightarrow \pi \eta] = 0.88\pm 0.23$, and 
$\Gamma [a_0(1450)\rightarrow \pi\eta' ]/\Gamma [a_0(1450)\rightarrow
\pi \eta] = 0.35 \pm 0.16$.
}
\label{Fig_a0_1450}
\end{figure}

\begin{table}
\begin{center}
$\begin{array}{|c|c|c|c|}
\hline
{\rm Fitted \hskip 0.2cm Parameters }  & {\rm{Eq.}} \hskip 0.2cm (\ref{SU3Lag}) &
{\rm{Eq.}} \hskip 0.2cm (\ref{int_simp})  & {\rm{Eq.}} \hskip 0.2cm (\ref{int_Lag})
\\ \hline 
A ({\rm GeV}^{-1}) & 0 &  0.10 \pm 0.12 &   1.03 \pm 0.12
\\ \hline
A'({\rm GeV}^{-1}) & -2.55 \pm 0.06 &  -2.60 \pm 0.06 & -3.53\pm0.12
\\ \hline
C ({\rm GeV}^{-1})  &   0    &  0  & 1.36\pm 0.27
\\ \hline
C' ({\rm GeV}^{-1}) &   2 A'    &  2 A'  & -6.56 \pm 0.27
\\ \hline
\multicolumn{4}{|c|}{
 {\rm  Predicted  \hskip 0.2cm Decay \hskip 0.2cm Widths }}
\\ \hline 
\Gamma^{tot} [a_0(1450)] ({\rm MeV}) & 265  &  265 & 265
\\ \hline
\Gamma [a_0(1450)\rightarrow K{\bar K}]/
\Gamma [a_0(1450)\rightarrow\pi\eta] & 0.55  & 0.53 & 0.53
\\ \hline
\Gamma [a_0(1450)\rightarrow \pi\eta']/
\Gamma [a_0(1450)\rightarrow\pi\eta] & 0.16  & 0.18 & 0.18
\\ \hline
\chi^2 &   1.161  &  1.157  & 1.157
\\ \hline
\end{array}$
\end{center}
\caption
{Best numerical values for the free parameters in the
scalar pseudoscalar pseudoscalar interaction Lagrangian, found by fitting
the prediction of our model for the total decay width and ratio of the
partial decay widths of $a_0(1450)$ to the experimental data.  The first
column corresponds to an interaction natural for $q{\bar q}$, while in the
second column interaction terms natural for $qq{\bar q}{\bar q}$ are also
included.  In the third column the more general interaction Eq. (4.1) is
considered.} 
\label{T_a0_1450}
\end{table}

Next, we include the $K_0^*(1430)$ in the picture. We take experimental
values \cite{PDG} $\Gamma^{tot} = 287 \pm 10 \pm 21$, and  $\Gamma_{\pi
K}/\Gamma^{tot} = 93 \pm 10 \%$, and search for regions that give 
$\Gamma [K_0^*(1430)\rightarrow\pi K] \approx 267 \pm 50$ MeV.  These are
shown in Fig. \ref{all_widths} with two parallel strips in the north-west
to south-east direction.   We see in the figure that within our 
model (\ref{int_simp}) there are overlaps of regions in parameter space $AA'$
that explain most of the decay properties of both $a_0(1450)$ and
$K_0^*(1430)$.

Now that we can understand the decay widths of 
the heavier scalars, we
explore the
possibility of explaining the decay widths of the light scalars within the
same  theoretical setup.  We proceed by further exploring  the parameter
space $AA'$ in the limit $C=0$ and $C'=2 A'$,  	
for regions that  explain decay
properties of the lighter physical
nonet  members $a_0(980)$ and   $\kappa(900)$.   We search for regions
consistent
with  $\Gamma [a_0(980)\rightarrow\pi\eta] \approx 65 \pm 5$ MeV in
agreement
with experimental measurement in \cite{Teige} as well as the theoretical
estimate in  \cite{Fariborz}.   
We also search for regions consistent with
\footnote{This is a width corresponding to the numerator, rather than
denominator of a partial wave amplitude  as explained in \cite{Black1}.
} 
$\Gamma [\kappa(900)\rightarrow\pi K ]\approx 40 \pm 5$ MeV  
in agreement with theoretical estimates of
the properties of $\kappa(900)$ given in \cite{Black2}.
The result is also  shown in Fig. \ref{all_widths}, indicating that there
are
regions in the parameter space of our model ($A\approx \pm 1$ and
$A'\approx\mp 3$) that are approximately consistent with the
decay properties of
the   light scalars {\it{in addition}} to those of the heavy scalars.

We have given 
in Table \ref{width_fit} our best fits for $A$, $A'$, $C$ and $C'$, 
resulting from comparing our theoretical prediction to the
experimental data.  We have also displayed in the same table the
predicted decay widths.  
In the limit $C=0$ and $C'= 2 A'$ 
(column one),  the resulting decay widths have the right order of
magnitude, although some of them are not within the ranges allowed by
experiment.   The fit gives $\frac{{\rm{A}}^{\prime}}{{\rm{A}}}= -2.4$
and so, as expected from the discussion of Fig. \ref{widths_fig} the widths
$\Gamma \left[ a_0 \left(1450 \right) \rightarrow K \bar K \right]$ and
$\Gamma \left[ a_0 \left(1450 \right) \rightarrow \pi \eta \right]$ have
the wrong order.  Outside this limit (column two), we get
a better agreement with experiment (as $\chi^2$ of fit also indicates),
and except for the  ratio $\Gamma [a_0(1450)\rightarrow K{\bar K}]/
\Gamma [a_0(1450)\rightarrow\pi\eta]$, all other decay widths are within
their experimentally allowed ranges.   We notice that in the general
case (column two), $C'\approx 2 A'$,  which means 
that the decay interaction for
nonet $N'$ remains close to that natural for $q{\bar q}$.
We also notice that in this general case,  $C\ne
0$, which is expected from our previous results on decays of the low lying
light scalars\cite{Black2}.

\begin{figure}
\centering
\epsfig{file=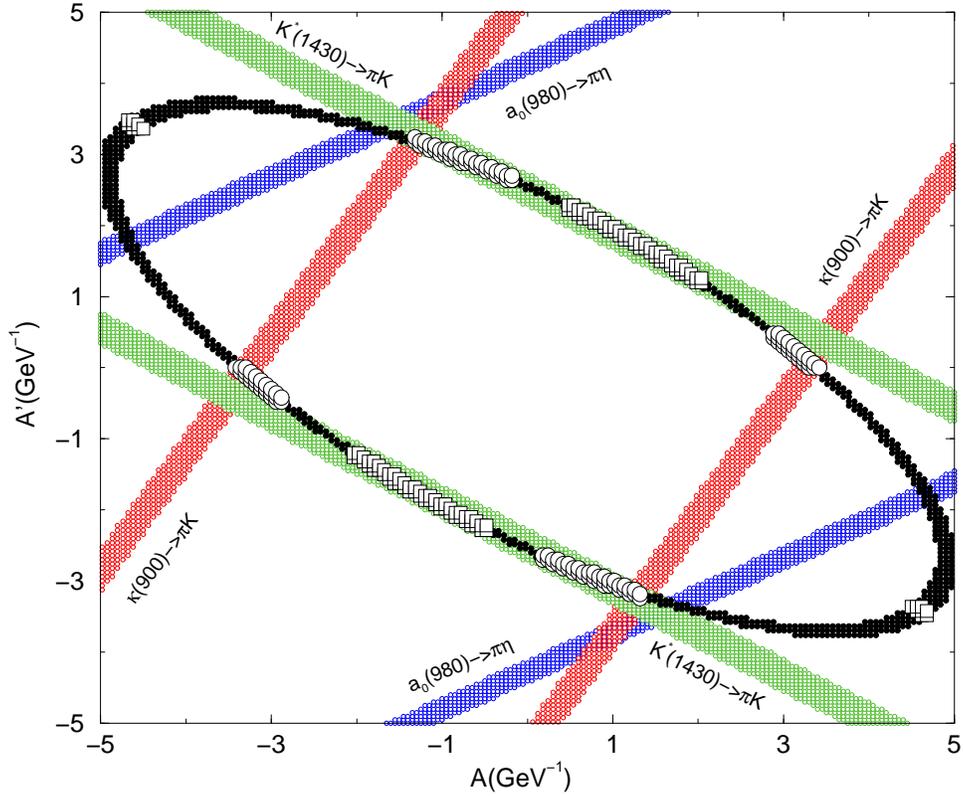, height=5in, angle=270}
\caption
{
Regions in the $AA'$ parameter space [see Eq. (\ref{int_simp})]
 consistent with the current
experimental and theoretical estimates on the decay widths of
$a_0(1450)$, $K^*_0(1430)$, $a_0(980)$ and $\kappa(900)$.   
Points on the ellipse are consistent with the total decay width of
$a_0(1450)$.  
Squares and circles respectively represent points
consistent with
the experimental ratio
$\Gamma [a_0(1450)\rightarrow K {\bar K} ]/\Gamma
[a_0(1450)\rightarrow \pi \eta] = 0.88\pm 0.23$, and
$\Gamma [a_0(1450)\rightarrow \pi\eta' ]/\Gamma [a_0(1450)\rightarrow
\pi \eta] = 0.35 \pm 0.16$.
}
\label{all_widths}
\end{figure}

\begin{table}
\begin{center}
$\begin{array}{|c|c|c|}
\hline
{\rm Fitted \hskip 0.2cm Parameters }  &  {\rm{Eq.}} \hskip 0.2cm
(\ref{int_simp}) & {\rm{Eq.}} \hskip 0.2cm (\ref{int_Lag})
\\ \hline 
A ({\rm GeV}^{-1}) & 1.40 \pm 0.12 &  1.19 \pm 0.16 
\\ \hline
A'({\rm GeV}^{-1}) & -3.26 \pm 0.07 &  -3.37 \pm 0.16
\\ \hline
C ({\rm GeV}^{-1})  &   0    &   1.05 \pm 0.49
\\ \hline
C' ({\rm GeV}^{-1}) &   2 A'    &  -6.87 \pm 0.50
\\ \hline
\multicolumn{3}{|c|}{
 {\rm  Predicted  \hskip 0.2cm Decay \hskip 0.2cm Widths }}
\\ \hline 
\Gamma^{tot} [a_0(1450)] ({\rm MeV}) & 274  &  263
\\ \hline
\Gamma [a_0(1450)\rightarrow K{\bar K}]/
\Gamma [a_0(1450)\rightarrow\pi\eta] & 0.30  & 0.42
\\ \hline
\Gamma [a_0(1450)\rightarrow \pi\eta']/
\Gamma [a_0(1450)\rightarrow\pi\eta] & 0.52  & 0.32
\\ \hline
\Gamma [K^*(1430)\rightarrow\pi K]  ({\rm MeV})& 245 & 298
\\ \hline
\Gamma [a_0(980)\rightarrow\pi\eta]  ({\rm MeV})& 57  & 65
\\ \hline
\Gamma [\kappa(900)\rightarrow\pi K]  ({\rm MeV})& 45   & 41
\\ \hline
\Gamma^{tot} [a_0(1450)]/\Gamma [K^*(1430) \rightarrow \pi K]& &\\
({\rm expected \hskip 0.2cm value:}\hskip 0.2cm 0.99\pm 0.24) 
& 1.12  & 0.88
\\ \hline
\chi^2 & 1.864 & 0.757 
\\ \hline
\end{array}$
\end{center}
\caption
{Best numerical values for the free parameters in the
scalar-pseudoscalar-pseudoscalar interaction Lagrangian, found by fitting
the prediction of our model 
for both the low lying and next-highest scalars
to the experimental data.  The first
and second columns correspond respectively to the limit ($C=0$ and $C'=2
A'$) and to the general case outside this limit.
}
\label{width_fit}
\end{table}

\section{Discussion}

  (i) We studied the properties of the $a_0(1450)$ and $K_0^*(1430)$ scalar
mesons (which are usually considered to belong to a conventional p-wave $q
\bar q$ nonet in the quark model) in a framework where a lighter scalar
nonet (of $ qq \bar q \bar q$ type) was also present.  It was found that
certain puzzling features of these two particles could be naturally
explained if the $q \bar q$ and $qq \bar q \bar q$ nonets mix with each
other to form new physical states.  The essential mechanism is driven
simply by the fact that the isospinor is lighter than the isovector in the
unmixed $qq \bar q \bar q$ multiplet.  

 (ii) Although we carried out the analysis in a $qq \bar q \bar q$ picture for
the unmixed light scalar nonet, it seems reasonable that it could also be
done for other models of the light scalars (like the unitarized quark model
\cite{Torn95,van86}, or molecular models \cite{Isgur}) in which they have somewhat different four-quark interpretations.

(iii) We did not investigate the heavier isoscalar particles because the
experimental situation is still rather ambiguous.  Clearly this is an
interesting future project.  

 (iv) In our treatment we used the simplest mixing term (\ref{mixing_mass}) and
obtained fairly good agreement with experiment.  The model can easily be
generalized to include different mixing terms in the effective Lagrangian;
for example
\begin{eqnarray}
{\rm{Tr}}\left[ {\cal{M}}\left( {N}N^{\prime} +
N^{\prime}{N} \right) \right]&,& \quad {\rm{Tr}}\left(
{N}\right) {\rm{Tr}}\left( N^{\prime} \right), \\ \nonumber
{\rm{Tr}}\left( {\cal{M}}{N} \right) {\rm{Tr}} \left(
N^{\prime} \right)&,& \quad {\rm{Tr}}\left( {\cal{M}}N^{\prime} \right) {\rm{Tr}} \left(
{N} \right).
\end{eqnarray}

  (v) Although our focus in this paper was on the heavier scalars, the model of
course describes the lighter ones too.  If we want to describe only
the lighter scalars, as in \cite{Black2,Fariborz}, we can imagine ``integrating
out'' the heavier scalars.  In the simplest approximation, based on
neglecting the symmetry breaking terms in Eqs. (\ref{mass-Lag-prime}) and (\ref{mass-Lag}), we would just replace
\begin{equation}
N^{\prime} \longrightarrow - \frac{\gamma}{2a^{\prime}}{N}.
\end{equation}
To check the consistency of this with our previous work we might ask
how much the decay coefficient of the light scalars [A in
Eq. (\ref{multiquarkLag})] gets modified due to this replacement.  Using
(\ref{SU3Lag}) and the identity in footnote 3 then gives 
\begin{equation}
{\rm{A}} \longrightarrow {\rm{A}} -
\frac{\gamma}{2a^{\prime}}{\rm{A}}^{\prime}.
\end{equation}
Using ${\rm{A}}= 1.2$ ${\rm{GeV}}^{-1}$ and ${\rm{A}}^{\prime} = -3.4$
${\rm{GeV}}^{-1}$ from column 2 of Table II, together with ${\gamma}^{2} =
0.33$ ${\rm{GeV}}^4$ and $a^{\prime} = 0.76$ ${\rm{GeV}}^2$ from (\ref{mass-Lag}) shows that ${\rm{A}}=
1.2$ ${\rm{GeV}}^{-1}$ in the present paper is to be replaced by ${\rm{A}}=
2.5$ ${\rm{GeV}}^{-1}$ in a model where the heavy scalars have been
eliminated.  This is in rough agreement with ${\rm{A}}=2.9$
${\rm{GeV}}^{-1}$ found in Table III of \cite{Black2}

\acknowledgments
We would like to thank Francesco Sannino for helpful discussions.  The work of D.B., A.H.F. and J.S. has been 
supported in part by the US DOE under contract
DE-FG-02-85ER 40231.

\end{document}